\documentclass[preprint,journal]{vgtc}            

\onlineid{1507}

\vgtccategory{Research}

\vgtcpapertype{Data Transformations}

\title{Class-constrained t-SNE: Combining Data Features and Class Probabilities}

\author{%
  \authororcid{Linhao Meng}{0000-0002-5611-5929}, 
  \authororcid{Stef van den Elzen}{0000-0003-1245-0503}, 
  \authororcid{Nicola Pezzotti}{0000-0001-7346-9182}, and 
  \authororcid{Anna Vilanova}{0000-0002-1034-737X}
}

\authorfooter{
  \item
  	Linhao Meng, Stef van den Elzen, Nicola Pezzotti, and Anna Vilanova are with Eindhoven University of Technology.
  	E-mail: \{l.meng1, s.j.v.d.elzen, n.pezzotti, a.vilanova\}@tue.nl

}

\abstract{%
  Data features and class probabilities are two main perspectives when, e.g., evaluating model results and identifying problematic items. Class probabilities represent the likelihood that each instance belongs to a particular class, which can be produced by probabilistic classifiers or even human labeling with uncertainty. Since both perspectives are multi-dimensional data, dimensionality reduction (DR) techniques are commonly used to extract informative characteristics from them. However, existing methods either focus solely on the data feature perspective or rely on class probability estimates to guide the DR process. In contrast to previous work where separate views are linked to conduct the analysis, we propose a novel approach, class-constrained t-SNE, that combines data features and class probabilities in the same DR result. Specifically, we combine them by balancing two corresponding components in a cost function to optimize the positions of data points and iconic representation of classes -- class landmarks. Furthermore, an interactive user-adjustable parameter balances these two components so that users can focus on the weighted perspectives of interest and also empowers a smooth visual transition between varying perspectives to preserve the mental map. We illustrate its application potential in model evaluation and visual-interactive labeling. A comparative analysis is performed to evaluate the DR results. 
}


\graphicspath{{figs/}{figures/}{pictures/}{images/}{./}} 

\usepackage{tabu}                      
\usepackage{booktabs}                  
\usepackage{lipsum}                    
\usepackage{mwe}                       

\usepackage{mathptmx}                  

\usepackage{amsmath}
\usepackage{amsfonts}
\usepackage{algorithm2e}
\usepackage{csquotes}

\newlength\mylen

\begin{document}
\definecolor{myblue}{HTML}{0682a7}
\definecolor{myred}{RGB}{0,0,0}


\firstsection{Introduction}

\maketitle

\firstsection{Introduction}

\maketitle

Classifiers are the most widely used machine learning models in fields such as finance, business, and healthcare. Developers make efforts to analyze classifiers to understand model behavior and identify potential problems during model development. Common model-agnostic strategies for examining classifiers involve analyzing data features, classification results, and the relationship between the two. Both data are bound to the same individual instances, implying that each instance encompasses its feature values and corresponding classification result. The classification result generally refers to the class label assigned to the instance. Of particular focus in our work are the class probabilities conveying the likelihood of instances belonging to each class. Class probabilities provide a more detailed evaluation of the classifier's performance than simple labels or summary performance measures and enable users to assess model behavior~\cite{squares}.

Since both data features and class probabilities are multi-dimensional, multi-dimensional data visualization techniques such as parallel coordinate plot, scatterplot matrix, or dimensionality reduction (DR) can be applied. Typically, data features and class probabilities are visualized in separate views connected by brushing and linking techniques. User interaction is often required to gain a complete understanding of their relationship. While it is possible to encode information from the other perspective as different visual channels in a single view~\cite{modelquality}, this approach may only capture a limited aspect of the other high-dimensional space. As each channel only encodes one variable or aspect of the high dimensional space, it can not provide a complete and accurate representation of the relationship between the two spaces.

Similar problems to classifier analysis can be found in document analysis where topic modeling results are typically visualized and analyzed with data features~\cite{TopicLens, VISTopic}. Topic modeling is a technique that discovers the underlying topics within a collection of documents and represents each document as a weighted combination of these topics. Probabilistic methods are often used, producing non-negative weights representing the probabilities of a document belonging to each related topic. Such probability data can also come from user labeling with uncertainty. Therefore, the problem in our study can be defined as analyzing two high-dimensional spaces of the same elements, where one space is the data features, and the other represents the probabilities of belonging to certain entities or classes.

\begin{figure}[tb]
  \centering 
  \includegraphics[width=0.98\linewidth]{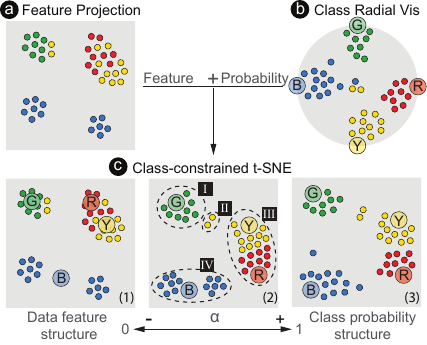}\vspace{-0.2cm}
  \caption{%
    A demonstration of DR results based on a single or combined perspective of data feature and class probability where data points are colored by predicted classes. (a) Data feature projection shows data feature characteristics by applying standard distance-based DR to data feature values. (b) Class radial visualization is employed to reveal probability-based model performance. (c) Class-constrained t-SNE enables the combination of data features and class probabilities.
  }\vspace{-0.4cm}
  \label{fig:demo}
\end{figure}

In contrast to previous work where multiple views are linked, in this work, we aim to merge both perspectives (i.e., data features and class probabilities) such that we can leverage both perspectives in one view. Both data feature structure and class probability structure are combined and displayed in a position-encoded manner. The data feature structure refers to instance similarity in data feature values. It is frequently visualized using general DR methods as shown in \cref{fig:demo}a. For the class probability structure, the between-instance similarities based on class probabilities matter, but also the relationship between instances and the classes. As a result, general DR methods that only preserve between-instance similarity are insufficient. 
Seifert and Lex have proposed the class radial visualization~\cite{radialclass} designed to display the relationship between instances and classes. The iconic representations of classes are displayed equally distributed around the perimeter of a circle. Instances are placed inside the circle based on their class probabilities, as depicted in \cref{fig:demo}b. The order and distribution of the iconic representations of classes have a significant influence on the visualization result and its interpretation. For example, it is not always clear among which classes the model is confused. In \cref{fig:demo}b, for data points in the center, it is unclear whether the model is confused between classes G and Y or among all classes. This ambiguity problem becomes more severe as the number of classes increases. 

Drawing inspiration from above projection methods, we use position proximity between data points to express data feature similarity. Position proximity between data points and iconic representation of classes named class landmarks conveys class probabilities. The ultimate placement of data points is determined by the tradeoff between preserving the data feature structure (\cref{fig:demo}c-1) or the class probability structure (\cref{fig:demo}c-3). By striking a balance between these two structures, we attain a merged projection result (\cref{fig:demo}c-2) that reveals patterns derived from the combination of data feature and class probability perspectives. Additionally, our method relieves the ambiguity problem by optimizing the positions of class landmarks in 2D space such that well-separated classes are drawn apart to mitigate data point overlap. As a result, the position proximity of class landmarks reflects class confusion.

The combination of data feature structure and class probability structure yields several advantages: (1) The merged projection result enhances visual cluster interpretability with class information compared to the one solely based on data feature values. The significant data feature structure manifested as every main visual cluster is preserved. Within each cluster, for example, in \cref{fig:demo}c-\mbox{I}, data points are arranged so that those with higher probabilities are more likely to be closer to the corresponding class landmark. Furthermore, clusters with the same predicted class are pulled towards the class landmark, as depicted in \cref{fig:demo}c-\mbox{IV}. (2) Incorporating class probability information improves the group separation inside visual clusters, where instances are difficult to distinguish solely based on data feature structure, as demonstrated in \cref{fig:demo}c-\mbox{III}. This impact is reminiscent of previous work using label information to improve the visual quality of projection results~\cite{featureextension,catsne}. However, our method goes beyond simply placing instances of the same class clustered together by leveraging class probability information and direct forces between class landmarks and data points to drag the most uncertain instances toward the borderline between different class groups. (3) The outliers, mainly a small number of instances unrepresentative of both perspectives or whose data feature similarities and class memberships conflict (e.g., \cref{fig:demo}c-\mbox{II}), are pushed out from the main visual clusters. These instances generally are worthy of further examination, such as in model analysis.

The main contribution of this work lies in a dimensionality reduction-based method, i.e., class-constrained t-SNE (as demonstrated in \cref{fig:demo}c), which has been implemented by imposing constraints upon the t-Distributed Stochastic Neighbor Embedding (t-SNE) algorithm. This method aims to combine and compare data feature structure and class probability structure to reveal the relationship between data features and class probabilities. Since both structures are displayed using position encoding, we can combine them by balancing two corresponding components in a cost function to optimize the positions of instances and class landmarks. Although coarse structures of both perspectives are considered by minimizing the cost function, it induces unavoidable interference when combining them. To alleviate this issue, we introduce an interactive user-adjustable parameter $\alpha$ to balance the two structures so that users can focus on the weighted perspectives of interest.
Meanwhile, our method empowers a smooth continuous visual transition between varying perspectives to preserve the mental map. It assists in tracing instances and facilitates the comparison among different projection results. We validate our method using several datasets and demonstrate its effectiveness in classifier performance and document topic analysis. \textcolor{myred}{We conduct a qualitative and quantitative comparative analysis with a baseline method to evaluate its superiority.} Additionally, we showcase its potential application in a visual-interactive labeling scenario, where adapting the structure balance parameter during model evolution enhances instance selection and labeling.

\section{Background}\label{background}
This section provides a brief introduction to the t-SNE algorithm, which is necessary to comprehend our approach.

t-SNE, as a nonlinear dimensionality reduction algorithm, is widely used for high-dimensional data visualization. It functions by preserving the neighborhood relationships between data points in the high-dimensional space when mapping them to a lower-dimensional space. To achieve that, t-SNE transforms the distance matrix between all data points in the high-dimensional space into a symmetric joint probability distribution $P$. Likewise, a joint probability distribution $Q$ is computed based on distances between the corresponding low-dimensional points. $P, Q \in \mathbb{R}^{n \times n}$ measure the pairwise similarities of data points in the high and low dimensional space, where $n$ is the number of data points. The goal is to let $Q$ match $P$ by iteratively optimizing the positions of low-dimensional points. This optimization process is achieved by minimizing the cost function $C$ that
measures the difference between $P$ and $Q$ using Kullback--Leibler (KL) divergence:
\begin{equation}\label{equ:KLtsne}
C = KL(P||Q) = \sum_{i}^{n}\sum_{j, j \neq i}^{n} p_{ij} \ln{\dfrac{p_{ij}}{q_{ij}}}.
\end{equation}
Here $p_{ij}$ indicates the similarity of data points $x_{i}$ and $x_{j}$ in the high-dimensional space. $p_{ij}$ is derived from the conditional probability $p_{j|i}$ and $p_{i|j}$. $p_{j|i}$ represents the probability that $x_{j}$ is picked as neighbor of $x_{i}$ based on a Gaussian probability density function centered at $x_{i}$ with variance $\sigma_{i}$. The variance $\sigma_{i}$ is determined by the perplexity parameter, which is the effective number of nearest neighbors considered when defining the neighborhood Gaussian probability density per point $x_{i}$. Mathematically, $p_{ij}$ is defined as:
\begin{equation}\label{equ:pd}
p_{ij} = \dfrac{p_{j|i}+p_{i|j}}{2n},  \quad
where \ p_{j|i}  = \dfrac{exp(-||x_{i}-x_{j}||^2/2\sigma_{i}^2)}{\sum_{k \neq i} exp(-||x_{i}-x_{k}||^2/2\sigma_{i}^2)}.
\end{equation}

In the low-dimensional space, a Student t-distribution with a single degree of freedom is used to convert distances into probabilities. Given the corresponding low-dimensional points $y_{i}$ and $y_{j}$, $q_{ij}$ describes the similarity of them given by 
\begin{equation}\label{equ:qd}
q_{ij} = \dfrac{(1+||y_{i}-y_{j}||^2)^{-1}}{\sum_{k \neq l} (1+||y_{k}-y_{l}||^2)^{-1}}.
\end{equation}

Details on t-SNE can be found in the original paper by Van der Maaten and Hinton~\cite{tsne} or other relevant sources~\cite{tsneParametric, attraction}. In our work, we employ the original t-SNE algorithm to reveal data feature structure. Simultaneously, we augment the cost function by introducing another component to incorporate class probability structure. Balancing these two components enables the combination of both structures. Details of our method are discussed in \cref{method}.

\section{Related Work}
In this section, we review previous work about visualization solutions for data features or class probabilities separately, followed by visualization methods concerning the combination. Finally, existing works regarding adding constraints into t-SNE are discussed in \cref{DR}. 

\subsection{Visualizing Data Feature or Class Probability}

Data feature vectors are commonly multi-dimensional data. Multiclass probabilities are multi-dimensional data as well, with each class considered one dimension. Popular multi-dimensional data visualization methods use parallel coordinate plots (PCP) and scatterplot matrices (SPLOM). Some work~\cite{iVisClassifier,multivariatelabel} apply PCP to show pairwise feature correlations. However, it is difficult to discern instance distribution for each class when applying PCP to class probabilities. To alleviate this issue, Chae et al.~\cite{Chae2017VisualizationFC} and Ren et al.~\cite{squares} align binning boxes with parallel coordinates to enrich the class-level performance information. 
Another standard method to deal with multi-dimensional data is to allow interactive selection of displayed dimensions in scatterplots~\cite{facets,datamodelspace}.

While the techniques mentioned above visualize the values for each dimension and can display pairwise dimension relationships, it remains challenging to infer the relationships between instances across multiple dimensions. Conversely, DR-based visualization methods project instances onto a 2D space and generate a layout where similarities between instances derived from multiple dimensions can be identified directly based on their spatial proximity~\cite{projection}. Therefore, DR-based visualization methods are widely used to present overall data distribution and similarity characteristics of data feature values in model analysis tasks~\cite{DFSeer, Vulnerabilities, modelwise}. However, to reveal class probability structure, it is important to indicate how each instance relates to each class. DR methods that only consider preserving the distances or similarities between pairwise instances based on class probabilities do not comply. Benefiting from a radial spring-based projection mechanism, RadViz visualization~\cite{RadViz} shows advantages in revealing relationships between instances and dimensions by the proximity of data points to the dimension anchors. Inspired by it, Seifert and Lex~\cite{radialclass,radialclassapplication} develop a class radial visualization for analyzing classification results and further apply it in interactive labeling tasks. However, as discussed before, the ambiguity problem remains and exacerbates with more class dimensions. \textcolor{myred}{In contrast, this problem can be alleviated by modifying the position of class anchors during the projection process~\cite{PE,DR4topic}}.


\subsection{Combining Data Feature and Class Probability}

Existing visualization works commonly utilize separate views and link them through interactions to facilitate analysis from both data feature and class probability perspectives~\cite{boxer, video}. Most offer fine-grained analysis at the instance or subset level from the selected primary perspective while displaying aggregated statistics from the other perspective~\cite{probwheel, manifold}. Some works support fine-grained analysis and instance selection from both perspectives.
One conventional method involves using a projection view that indicates data feature structure alongside a scatterplot where one axis encodes class probabilities~\cite{topicmodeling, documentretrieval}. However, with only two axes of a scatterplot, it can encode the class probabilities of at most two classes, which is insufficient for a multiclass scenario. Separate views require user interactions to examine the same instances from different perspectives of data features and class probabilities so that users can gain an overall understanding of instance attribution in data feature structure and class probability space.


There are several previous works aiming to achieve the combination of two perspectives in a single view. ModelTracker~\cite{modeltracker} aligns instance squares on a horizontal axis according to prediction scores and connects the nearest neighbors in the feature space by lines when hovering over an instance square. It does not provide an overview of data feature structure and cannot easily be extended to multi-class cases. Schneider et al.~\cite{datamodelspace} use the X axis to encode classification probability and the Y axis to encode the attribute value of one selected feature or instance similarity in a scatterplot. Since this approach only considers the biggest probability prediction for each instance, it is incapable of revealing class confusion derived from probability information among all classes. Instead of using class probabilities, some work~\cite{NJtree,UTOPIAN} has incorporated class membership information directly into data feature similarity by perturbing the data feature distance matrix. They build a neighbor-joining tree or generate DR results based on whether or not instances belong to the same class. However, these methods only consider the class consistency of the instances but not the relationship between instances and specific classes reflected by class probabilities. Our work explores the combination of the data feature and class probability perspectives in one view. We use dimensionality reduction as the backbone of the visualization, which to the best of our knowledge, has yet to be explored in previous work.

\subsection{Integrating Constraints into t-SNE} \label{DR}

The major objectives of integrating constraints into t-SNE in existing work are (1) improving the stability of projection results, \textcolor{myred}{(2) incorporating additional information to enhance embedding quality, and (3) accommodating specific layout requirements.} The first objective is commonly achieved by the fixed point constraint. It can be done by fixing the positions of some anchor points and performing iterative optimization of other points~\cite{TopicLens, PIVE, liu1}. This method is typically used to maintain view consistency when the user interacts with a projection view. Another version of the fixed point constraint is to add an extra loss term to generate consistent projections for the temporal evolution of the same observations~\cite{dynamictSNE}. The loss restricts relative positions between the identical points of adjacent time frames. \textcolor{myred}{To accomplish the second objective, additional information, such as class labels, is utilized to enforce proximity constraints among data points, bringing points of the same class closer and increasing the separation between points of different classes. Constraints can be applied during the computation of the (dis)similarity matrix~\cite{catsne,supervisedtSNE} or included in the cost function~\cite{supervisedDR, HCtSNE}. Likewise, in the domain of topic modeling,  TopicLens~\cite{TopicLens} and UTOPIAN~\cite{UTOPIAN} enhance the visual clarity of topic modeling results by adjusting pairwise distances between document representations in accordance with their membership in topic clusters when using t-SNE. }

\textcolor{myred}{Our goal extends beyond utilizing class labels to enhance class separation among data points. Instead, we aim to impose constraints on the proximity between data points and class landmarks, which accords with the third objective. The "landmark" concept can be found in some previous work~\cite{landmark2003,landmark2004} where a subset of the data is chosen as "landmark points" to guide the projection of other data points. However, in our work, each class landmark represents one dimension of the data in terms of class probabilities.}
\textcolor{myred}{A typical example of meeting specific layout requirements can be found in Liu et al.'s work~\cite{crowdsource}. By introducing class constraint arcs as class representations in a circular-based layout, instance uncertainty (crowdsourced label consistency) is reflected by relative positions between instance points and class arcs. Meanwhile, the most uncertain ones are pushed into the circular center. To achieve this layout, an additional component is added to the original t-SNE cost function, which transforms the constraint of revealing crowdsourced label information into another KL function.} Nevertheless, since positions of class constraint arcs are fixed in the circular layout, the projection result has a similar ambiguity issue as in RadViz. Additionally, it cannot directly present which classes are more likely to be confused. Our approach also employs weighted KL divergences in the cost function to combine and balance different components. However, rather than relying on static class arcs, we use iconic representations -- class landmarks -- whose positions are concurrently optimized with data points. This innovative technique alleviates the shortcomings associated with fixed layouts.



\begin{table*}[tb]
  \caption{%
  Visual patterns when combining data features and class probabilities. %
  }\vspace{-0.2cm}
 \label{tab:cases}
   \scriptsize%
  \centering%
  \setlength{\tabulinesep}{0.15em}
  \begin{tabu}to \linewidth {c p{8cm}  X[1,L,m]}
    \tabucline [1pt]{-}
       &\begin{tabu}{X[c]X[0.2cm]X[c]X[0.1cm]X[c]}
    \textbf{Feature(\textcolor{myblue}{f})} & + & \textbf{Probability(\textcolor{myblue}{p})} & = & \textbf{Combination(\textcolor{myblue}{c})} \\
  \end{tabu} & \textbf{Description} \\
      \tabucline [1pt]{-}
      P1 &
          \parbox[p]{8cm}{\includegraphics[width=\linewidth]{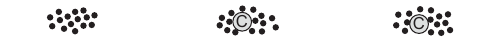}} & A cluster (\textcolor{myblue}{f}) with the same class C (\textcolor{myblue}{p}) retains in (\textcolor{myblue}{c}) \\
\tabucline [0.5pt]{-}
      P2 & \parbox[p]{8cm}{\includegraphics[width=\linewidth]{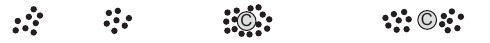}} &  Two clusters (\textcolor{myblue}{f}) with the same class C (\textcolor{myblue}{p}) get closer in (\textcolor{myblue}{c})  \\
\tabucline [0.5pt]{-}
      P3 & \parbox[p]{8cm}{\includegraphics[width=\linewidth]{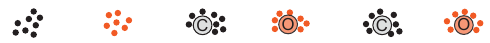}}  &  Two clusters (\textcolor{myblue}{f}) with high-probability classes C and O (\textcolor{myblue}{p}) retain in (\textcolor{myblue}{c})  \\
\tabucline [0.5pt]{-}
      P4 &  \parbox[p]{8cm}{\includegraphics[width=\linewidth]{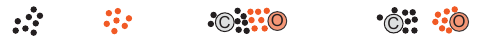}}  &  Two clusters (\textcolor{myblue}{f}) with classes C and O but with uncertain instances in between (\textcolor{myblue}{p}) get closer in (\textcolor{myblue}{c})  \\
\tabucline [0.5pt]{-}
      P5 & \parbox[p]{8cm}{\includegraphics[width=\linewidth]{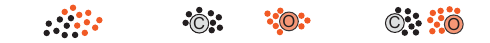}}  &  A cluster with low class mixing (\textcolor{myblue}{f}) and high-probability for classes C and O (\textcolor{myblue}{p}) get separated in (\textcolor{myblue}{c})  \\
\tabucline [0.5pt]{-}
      P6 &  \parbox[p]{8cm}{\includegraphics[width=\linewidth]{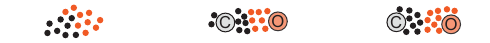}}  &  A cluster with low class mixing (\textcolor{myblue}{f}) and uncertain instances in between for classes C and O (\textcolor{myblue}{p}) slightly get separated in (\textcolor{myblue}{c})  \\
\tabucline [0.5pt]{-}
      P7 & \parbox[p]{8cm}{\includegraphics[width=\linewidth]{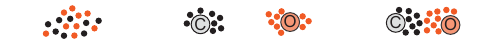}}  &  A cluster with high class mixing (\textcolor{myblue}{f}) and high-probability for classes C and O (\textcolor{myblue}{p}) generate borderline in (\textcolor{myblue}{c})  \\
\tabucline [0.5pt]{-}
      P8 &  \parbox[p]{8cm}{\includegraphics[width=\linewidth]{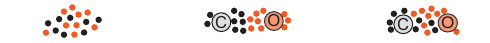}}  &  A cluster with high class mixing (\textcolor{myblue}{f}) and  uncertain instances in between for classes C and O (\textcolor{myblue}{p}) generate unclear borderline in (\textcolor{myblue}{c}) \\
\tabucline [0.5pt]{-}
      P9 & \parbox[p]{8cm}{\includegraphics[width=\linewidth]{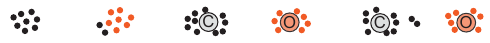}}  & Two clusters (\textcolor{myblue}{f}) primarily with classes C and O (\textcolor{myblue}{p}) retains in (\textcolor{myblue}{c}). Few instances classified as C (\textcolor{myblue}{p}) but in another cluster (\textcolor{myblue}{f}) are drawn towards the cluster of class C (\textcolor{myblue}{c}). \\
\tabucline [0.5pt]{-}
      P10 &  \parbox[p]{8cm}{\includegraphics[width=\linewidth]{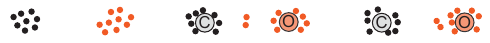}}  &  Two clusters (\textcolor{myblue}{f}) with classes C and O respectively (\textcolor{myblue}{p}) retains in (\textcolor{myblue}{c}). Few instances in the cluster of class O (\textcolor{myblue}{f}) and classified as O but with low probabilities (\textcolor{myblue}{p}) are slightly pushed away from the original cluster(\textcolor{myblue}{c}).  \\
     \tabucline [1pt]{-}
  \end{tabu}\vspace{-0.2cm}
\end{table*}

\section{Class-constrained t-SNE}\label{method}

Class-constrained t-SNE is a DR-based method that combines the data feature and class probability perspectives in the projection result. In this section, we start by clarifying the objectives and requirements of our method. Then we present the method itself, which adapts the original t-SNE method to display the data feature structure while incorporating class probabilities as constraints to reveal the class probability structure. 

\subsection{Goal Analysis} \label{sec: referencemethod}

DR methods reduce the dimensionality of the original data features and facilitate the visualization of salient structures, which we refer to as the data feature structure. Additional information such as labels, external features, or user feedback can adapt the low-dimensional representations to the users' needs by integrating constraints into DR methods~\cite{DRsurvey}. One way to incorporate class probabilities for the same instances is to treat them as additional features added to the original data feature space, which will directly perturb the pairwise instance similarity or distance matrix~\cite{NJtree, featureextension}. \textcolor{myred}{Likewise, we can incorporate a constraint into the original DR cost function to boost the clustering of data points with similar class probabilities~\cite{equivalenceconstrain}. A naive approach would be linearly combining two KL functions. The first KL function is the original t-SNE cost function for the data feature space. The second KL function is an additional one, designed in a similar manner but employing pairwise class probability similarity of data points as the input.} However, this method hides the correlation between data features and class probabilities. In addition, the relationship between instances and the classes themselves is ignored when converting class probabilities to the similarities among instances. Thus the class-level information that can be presented is limited.

As an alternative, class probabilities indicate similarities between instances and classes. We expect this relationship to be reflected in the DR results. Therefore, in this work, the \textit{goal} we aim at is combining and exhibiting two components in the projection result. One component is the data feature structure reflected by pairwise instance similarity based on data feature values. The other is the class probability structure derived from the relationship between instances and classes.

In the projection results using t-SNE, data feature structure is manifested by the clustering of data points that share similar data feature values. To better present class probability structure, we introduce the iconic representations of classes -- class landmarks as extra points in the low dimensional space. The use of class landmarks enables the identification of class probabilities based on the proximity of data points to these class landmarks. Furthermore, the proximity between the class landmarks indicates the confusions between the classes, as reflected in the class probabilities.

We expect that when combining these two structures, specific visual patterns (\textbf{P1-10}) will emerge, as summarized in \cref{tab:cases}. \textbf{P1} and \textbf{P2} refer to scenarios where one or more clusters share the same class. In the combined projection result, these clusters are retained around their corresponding class landmarks, while the positions of data points are adjusted according to their class probabilities. Specifically, instances with higher probabilities are relocated nearer to the class landmark. \textbf{P3} and \textbf{P4} illustrate situations where distinct visual clusters are assigned as different classes. In \textbf{P3}, the classes are well separated with high probabilities, while in \textbf{P4}, the classes are not well separated, with instances of relatively low probabilities interspersed. In \textbf{P5-8}, classes cannot be distinguished directly in the original data feature space, resulting in instances of mixed classes within a single visual cluster. When combined with class probability information, the inner structure of the cluster is expected to emerge, with borderlines generated between data points of different classes. Data points at the borderlines are the most confused between the classes. However, the degree to which data points of different classes are separated within the cluster depends on their degree of certainty in the class probability space and the degree of separation in the original data feature space. This is illustrated in \textbf{P5-7} and \textbf{P8}, which show how clear or unclear borderlines can result from these factors. \textbf{P9} show outliers whose data feature pattern and class result conflict. In this case, the outliers are pushed away from the original cluster. In \textbf{P10}, the data feature pattern appears to align with the class result. But a few instances within the cluster exhibit relatively low probabilities compared to the rest in that cluster. As a result, these instances are slightly pushed away from the original cluster. \textcolor{myred}{Note that the combination patterns, as observed, do not accurately reflect a single perspective. Therefore, a flexible transformation that transitions seamlessly from one perspective to another would be required.} In the following subsection, we propose the method we expect to produce the results defined above.

\subsection{Method}

Given a data set $X=\{x_{1},x_{2},...,x_{n}\}$ where $x_{i}$ is a multi-dimensional feature vector and $n$ is the number of instances, and a corresponding $m$-class probability result $T=\{(t_{11},t_{12},...,t_{1m}),...,(t_{n1},t_{n2},...,t_{nm})\}$ where $t_{iu}$ is the probability of $x_{i}$ belonging to class $u$, we aim to convert $X$ into two-dimensional data $Y=\{y_{1},y_{2},...,y_{n}\}$ displayed in a scatterplot which satisfies our \textit{goal} as defined above. 

To represent the data feature structure, we propose using the conventional t-SNE optimization approach to preserve neighborhoods and cluster structures.
To reveal class probabilities in the two-dimensional map, we introduce $m$ class landmarks in the two-dimensional space $V=\{v_{1},v_{2},...,v_{m}\}$ where each $v_{u}$ denotes the position of one class landmark, the iconic representation of class $u$. \textcolor{myred}{The high-dimensional representation of a class landmark in terms of class probabilities is one of the unit vectors $E=\{e_{1},e_{2},...,e_{m}\}$. Each unit vector represents a specific dimension in the m-dimensional space.} Position proximity between data points $Y$ and class landmarks $V$ reflects class probability information. Specifically, data points more certain to be one class are closer to the corresponding class landmark than the others. Data points confused by two or a few classes are drawn toward the borderlines between the corresponding class clusters. 

To achieve the above hypothesis, we introduce two cost functions, denoted as $fc_{1}$ and $fc_{2}$, for optimizing the projection results to represent the data feature structure and the class probability structure, respectively. These cost functions are specifically designed to capture the distinct characteristics of each structure given by
\begin{align}
      fc_{1}&=KL(P^{d}||Q^{d}), \label{eq:cost1} \\
     fc_{2}&= \dfrac{1}{n} \sum_{i=1}^{n} (KL(P^{c}_{i}||Q^{c}_{i}) + \lambda \cdot D), \quad D = \dfrac{1}{m}  \sum_{u=1}^{m} p^{c}_{iu} \cdot ||y_{i}-v_{u}||^2.
     \label{eq:cost2}
\end{align}

$fc_{1}$ is the same as \cref{equ:KLtsne} used in the original t-SNE method. $P^{d}, Q^{d} \in \mathbb{R}^{n \times n}$ are joint probability distributions that measure pairwise similarities of data points in the high-dimensional data feature space and low-dimensional space, which have been formulated in \cref{background}.

$fc_{2}$ enforces the position proximity between data points and class landmarks based on class probabilities. $P^{c}_{i}$ and $Q^{c}_{i}$ are m-dimensional vectors. $P^{c}_{i}$ measures the similarities of $x_{i}$ and $m$ classes in the high-dimensional space. $p^{c}_{iu} \in P^{c}_{i}$, the probability that $x_{i}$ belongs to class $u$, is drawn from the class probability result $T$ as
\begin{equation}\label{equ:pc}
p^{c}_{iu} \textcolor{myred}{= t_{i} \cdot e_{u}} = t_{iu}, \quad u=1,2,...,m.
\end{equation}
For the 2D counterparts $y_{i}$ and $V$ of $x_{i}$ and $m$ classes, we use a probability distribution $Q^{c}_{i}$ to model the similarities of data point $y_{i}$ and class landmarks $V$. We define the probability $q^{c}_{iu} \in Q^{c}_{i}$ measuring the similarity between data point $y_{i}$ and class landmark $v_{u}$ in the 2D map as
\begin{equation}\label{equ:qc}
    q^{c}_{iu}  = \dfrac{(1+||y_{i}-v_{u}||^2)^{-1}}{\sum_{s=1}^{m} (1+||y_{i}-v_{s}||^2)^{-1}}.
\end{equation}
We aim to find the 2D representations $y_{i}$ and $V$ of instance $x_{i}$ and classes, which minimizes the mismatch between $P^{c}_{i}$ and $Q^{c}_{i}$. Since $P^{c}_{i}$ and $Q^{c}_{i}$ are probability distributions, we also use KL divergence $KL(P^{c}_{i}||Q^{c}_{i})$ to measure the faithfulness that using $Q^{c}_{i}$ to model $P^{c}_{i}$. By minimizing the sum of KL divergences over all instances, we aim to obtain a layout of all data points and class landmarks, revealing the relationship between instances and classes.

Inspired by t-SNE, we use the heavy-tailed Student t-distribution to convert distances between data points and class landmarks to probabilities when calculating $Q^{c}$. It allows moderate class probability to be modeled by a relatively large distance, eliminating the unwanted force between data points and less likely class landmarks and giving more space between class landmarks to show the data points confused by certain classes in the map. To maintain consistency in the variance of the probability distribution used to calculate $Q^{d}$, we also use a single degree of freedom, \textcolor{myred}{as demonstrated to be appropriate for 2D embeddings~\cite{tsneParametric}, when calculating $Q^{c}$. It is worth noting that while the degree of freedom for t-SNE is adjustable, using the same setting for both $Q^{d}$ and $Q^{c}$ contributes to a fair and unbiased representation of the resulting embedding when combining the two structures.}

By minimizing $KL(P^{c}_{i}||Q^{c}_{i})$, we ensure the position proximity between data point $y_{i}$ and all class landmarks $V$ reflect class probability of $x_{i}$. However, in some cases, infinite possible $y_{i}$ could satisfy the minimization requirement. For example, for $x_{i}$ with an equal probability of 0.5 for two classes, any two-dimensional map $y_{i}$ ending up on the perpendicular bisector of the corresponding class landmarks satisfies this probability constraint. To relieve this issue, we add a distance penalty $D$ in $fc_{2}$ to ensure there is only one optimal value for $y_{i}$, which is the midpoint of the two class landmarks in the above example. As a result, we ensure a stable two-dimensional map can be attained as an optimal solution. This distance penalty is weighted by a parameter $\lambda > 0$. 

Note that $fc_{1}$ only constrains the relative distances between data points, while $fc_{2}$ constrains the relative distances between data points and class landmarks. $fc_{2}$ can be minimized by optimizing the positions of data points, class landmarks, or both. In other words, the positions of data points are affected by the optimization of both $fc_{1}$ and $fc_{2}$. The amount of influence of different perspectives depends on how much weight we put on them. In contrast, positions of class landmarks can always be optimized to minimize $fc_{2}$ no matter where data points are. Therefore, we define the cost functions $C_{d}$ for optimizing instance representations $Y$ and $C_{c}$ for optimizing class landmarks $V$ as
\begin{align}
     C_{d} &= (1-\alpha) \cdot fc_{1} + \alpha \cdot fc_{2}, \\
    C_{c} &= fc_{2}.
\end{align}
The positions of data points are a tradeoff between two objectives $fc_{1}$ and $fc_{2}$ balanced by the parameter $\alpha \in [0,1]$. We make this parameter configurable, enabling users to manipulate it and focus on the weighted perspectives of interest. 

\subsection{Optimization}\label{optimization}

\RestyleAlgo{ruled}

\begin{algorithm}[tb!]
\caption{Class-constrained t-SNE}\label{alg:cap}
\KwData{data set $X=\{x_{1},x_{2},...,x_{n}\}$, class probability $T=\{(t_{11},t_{12},...,t_{1m}),...,(t_{n1},t_{n2},...,t_{nm})\}$, structure balance parameter $\alpha$, distance penalty weight $\lambda$, perplexity $Perp$, number of iterations $K$, learning rate $\eta$, momentum $\mu(k)$.}
\KwResult{2D data representation $Y$, 2D class landmarks $V$.}
\SetKwBlock{Beginn}{beginn}{ende}
\Begin{
compute pairwise data affinities $P^{d}$ with $Perp$ (using \cref{equ:pd})\;
compute data class affinities $P^{c}$ (using \cref{equ:pc})\;
Initialize $Y$ and $V$ from $N(0,10^{-4}I)$\;
\For{k=1 \KwTo K}{
  compute pairwise 2D data affinities $Q^{d}$ (using \cref{equ:qd})\;
  compute 2D data class affinities $Q^{c}$ (using \cref{equ:qc})\;
  set $Y^{(k)}=Y^{(k-1)}+\eta\frac{\partial C_{d}}{\partial Y}+\mu(k)(Y^{(k-1)}-Y^{(k-2)})$\;
  set $V^{(k)}=V^{(k-1)}+\frac{\eta m}{n}\frac{\partial C_{c}}{\partial V}+\mu(k)(V^{(k-1)}-V^{(k-2)})$\;
}
}
\end{algorithm}

\begin{figure}[tbp]\vspace{-0.2cm}
  \centering
  \begin{subfigure}[b]{0.5\columnwidth}
  	\centering
  	\includegraphics[width=0.7\textwidth]{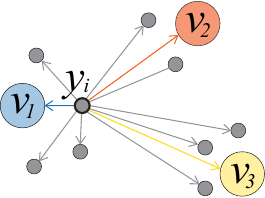}
  	\caption{Forces on data point $y_{i}$}
  	\label{fig:dataforce}
  \end{subfigure}%
  \hfill%
  \begin{subfigure}[b]{0.5\columnwidth}
  	\centering
  	\includegraphics[width=0.7\textwidth]{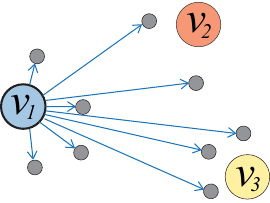}
  	\caption{Forces on class landmark $v_{1}$}
  	\label{fig:classforce}
  \end{subfigure}\vspace{-0.4cm}
  \subfigsCaption{Illustration of using force to interpret the gradient. Grey points denote data points, and each circle labeled as $v_{i}$ is one class landmark.}
  \label{fig:ex_subfigs}\vspace{-0.4cm}
\end{figure}

\begin{figure*}[tbh]
  \centering 
  \includegraphics[width=0.98\linewidth]{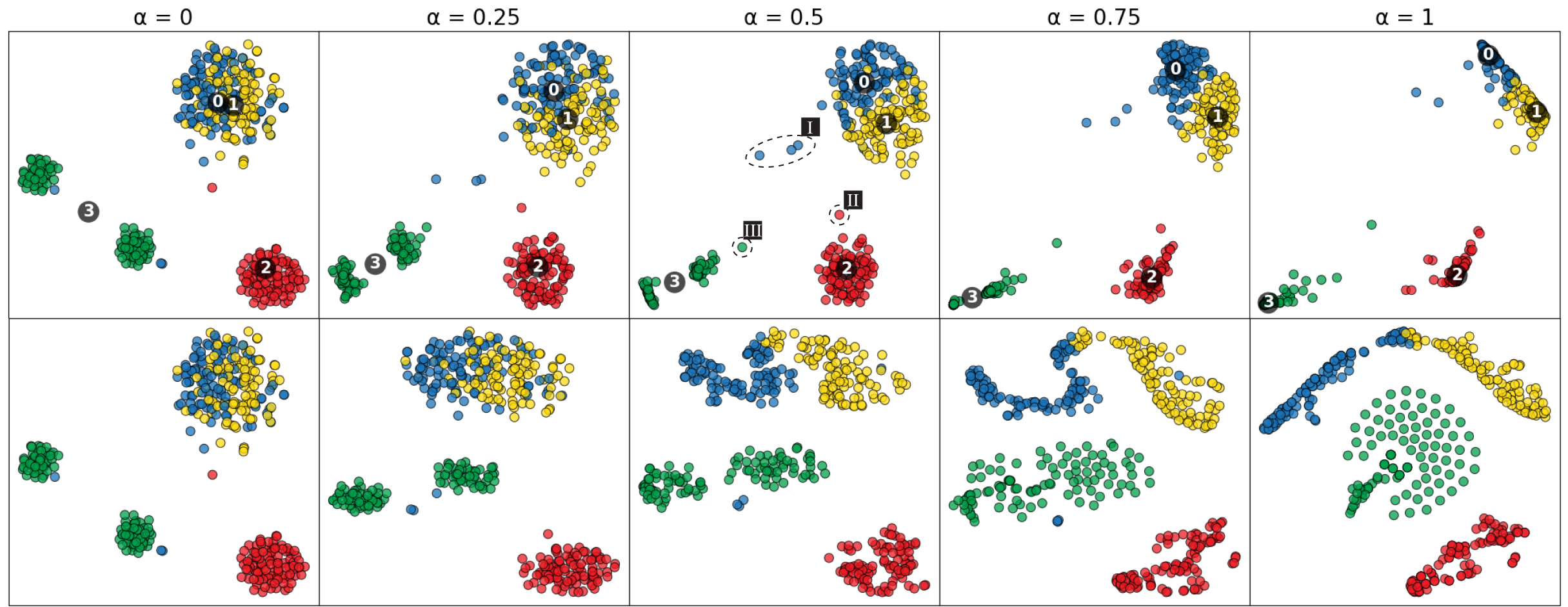}\vspace{-0.2cm}
  \caption{%
    Projection results of a synthetic dataset with various $\alpha$ values using our method (top) and a baseline method (bottom).
  }
  \label{fig:syntheticalpha}\vspace{-0.2cm}
\end{figure*}
\begin{figure*}[tbh]
  \centering 
  \includegraphics[width=0.98\linewidth]{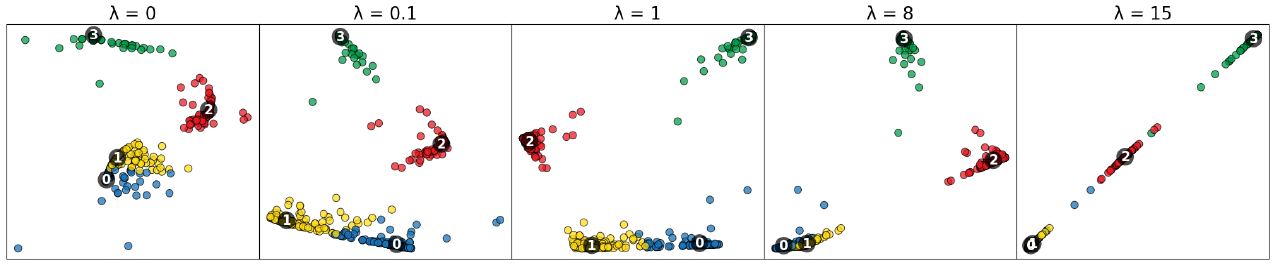}\vspace{-0.2cm}
  \caption{%
    Projection results of a synthetic dataset with various $\lambda$ values ($\alpha=1$). 
  }
  \label{fig:lambda}\vspace{-0.4cm}
\end{figure*}

We use a similar optimization method as proposed by Van der Maaten and Hinton~\cite{tsne}. The algorithm is described in \cref{alg:cap}. The minimization of cost functions is achieved by gradient descent. The gradients are computed as
\begin{align*}
\frac{\partial C_{d}}{\partial y_{i}} &= (1-\alpha) \cdot \frac{\partial fc_{1}}{\partial y_{i}}  + \alpha \cdot \frac{\partial fc_{2}}{\partial y_{i}}, \\
\frac{\partial fc_{1}}{\partial y_{i}} &=  4\sum_{j}(p^{d}_{ij}-q^{d}_{ij})(y_{i}-y_{j})Z_{ij}, \\
\frac{\partial fc_{2}}{\partial y_{i}} &= \frac{2}{n} \sum_{u}((p^{c}_{iu}-q^{c}_{iu})(y_{i}-v_{u})Z_{iu} + \frac{\lambda}{m} p^{c}_{iu} (y_{i}-v_{u}) ), \\
\frac{\partial C_{c}}{\partial v_{u}} &= \frac{2}{n} \sum_{j}((p^{c}_{ju}-q^{c}_{ju})(v_{u}-y_{j})Z_{ju} + \frac{\lambda}{m} p^{c}_{ju} (v_{u}-y_{j})), \\
where \ & Z_{ij}=(1+|| y_{i}-y_{j} ||^{2})^{-1} \ and \ Z_{iu} = (1+|| y_{i}-v_{u} ||^{2})^{-1}.
\end{align*}
$\partial C_{d} / \partial y_{i}$ can be interpreted as the weighted resultant force exerted on the data point $y_{i}$ by all other data points and class landmarks. As shown in \cref{fig:dataforce}, we illustrate these forces in grey forces from other data points ($ \partial fc_{1}/\partial y_{i}$) and class hue encoded forces from class landmarks ($\partial fc_{2}/\partial y_{i}$). Each individual force acts to repel or attract the point mainly based on their distance in the 2D space and the difference between the target probability and the probability derived from the 2D embeddings. Similarly, the gradient of $C_{c}$ with respect to the class landmark $v_{u}$ ($\partial C_{c}/\partial v_{u}$) can be interpreted as the force exerted on the class landmark $v_{u}$ by all data points, as exemplified in \cref{fig:classforce}. 

We simultaneously optimize the positions of 2D data points and class landmarks as a key feature of our method. By minimizing $C_{d}$ to optimize the positions of data points regarding class landmarks as fixed and $C_{c}$ to optimize the positions of class landmarks regarding data points as fixed, we can find the 2D representation of data points and class landmarks that preserves a weighted perspective of the data feature structure and class probability structure. Importantly, by optimizing class landmarks' position $V$, we alleviate the ambiguity caused by the fixed layout in the class radial visualization~\cite{radialclass}. Meanwhile, the final layout of class landmarks reflects class confusion. Furthermore, by gradually adjusting $\alpha$ and initializing the optimization process with the previous DR result \textcolor{myred}{and no early exaggeration}, we implement a smooth transition from displaying a single structure to combining both structures and then transitioning to displaying the other structure. It assists in preserving the mental map when tracing instances among various DR results and facilitates the comparison of different perspectives. 

The most computationally expensive part of the original t-SNE method is the computation of the pairwise similarities between all pairs of data points. It typically requires $O(n^2)$ operations, where n is the number of data points. Some studies have proposed alternative approaches that achieve a time complexity of $O(n logn)$~\cite{BHt-SNE}, or $O(n)$~\cite{lineartSNE}. Besides that, our method has the additional computation of the similarity between data points and class landmarks. It has a time complexity of $O(n m)$ where m is the number of classes. Given the limited number of classes, the computation efficiency of class-constrained t-SNE is primarily constrained by the size of the dataset.

\section{Experiments}

In this section, we present the results of multiple exemplary experiments to illustrate the functioning of our method. We investigate the influence on DR results of the value of structure balance parameter $\alpha$ and distance penalty weight $\lambda$. We utilize a synthetic dataset initially, followed by two real-world datasets -- one for classifier analysis and the other for document topic analysis. All data points in the figures are color-coded based on their highest class probabilities rather than true labels.


\begin{figure*}[tbh]
  \centering 
  \includegraphics[width=\linewidth]{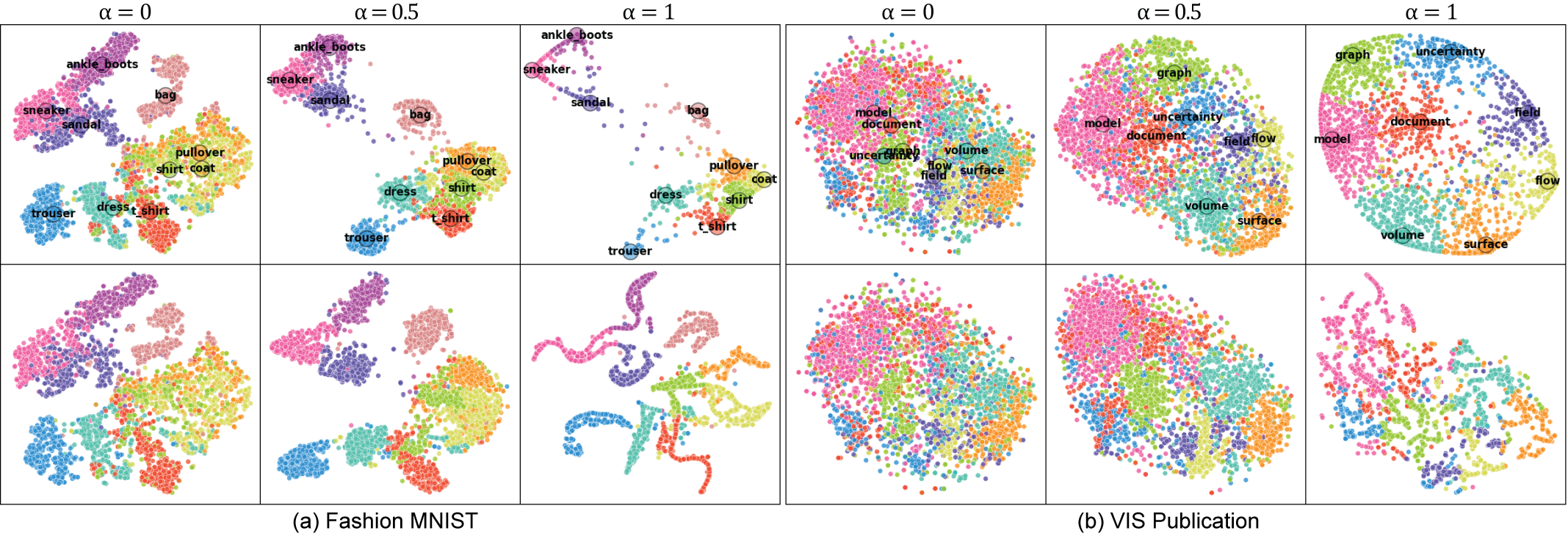}\vspace{-0.2cm}
  \caption{%
    Projection results of two datasets with three $\alpha$ values using our method (top) and a baseline method (bottom).
  }
  \label{fig:cases}\vspace{-0.4cm}
\end{figure*}

\subsection{A Synthetic Example}

We generate a synthetic dataset to validate our method and illustrate the produced visual patterns. This 10D dataset primarily comprises five Gaussian clusters labeled as four classes. Two clusters, each with 100 instances and very close centers, are labeled 0 and 1, respectively. Additionally, two clusters, each comprising 50 instances, are given the label 3, while the remaining cluster, containing 100 instances, is labeled 2. A few points are generated as noise points, which have similar vector values to clusters labeled 3 but are labeled 0. To train our model, we used 70\% of the data with a random forest model, which achieved an accuracy of 84\% on the test dataset. Subsequently, we applied the model to generate class probabilities for the entire dataset. 

\Cref{fig:syntheticalpha} (top) shows the projection results using class-constrained t-SNE with five $\alpha$ values ranging from 0 to 1 with step 0.25 and $\lambda=0.5$. All the projection results were obtained by initializing the algorithm with the preceding projection result, except for the first one of $\alpha=0$ with random initialization. This helps to provide a smooth transition between continuous changes of $\alpha$. The projection result of $\alpha=0$ reveals the data feature structure exclusively. The observed cluster patterns largely align with our data generation process. The projection result of $\alpha=1$ shows the class probability structure. Four clusters correspond to four classes, each guided by the corresponding class landmark. The position proximity of class landmarks reflects class confusion. Classes 2 and 3 are distinguishable from other classes, while classes 0 and 1 are relatively confused with each other. Data points are placed based on their class probabilities, in which those with high probabilities are closer to their corresponding class landmarks. Plenty of data points between class landmarks 0 and 1 explain the class confusion between classes 0 and 1. When $\alpha$ equals 0.5, the combined visual patterns emerge. The clusters primarily associated with classes 2 and 3 can be attributed to \textbf{P1} and \textbf{P2} in \cref{tab:cases}, respectively. The ambiguous boundary between classes 0 and 1 is consistent with \textbf{P8}. The identification of outliers denoted as \mbox{I}, which are located within clusters associated with class 3 in the data feature structure but classified as class 0, is indicative of \textbf{P9}. Similarly, the presence of the outlier \mbox{II}, which is not representative of class 2 in the data feature structure, is another example of \textbf{P9}. The outlier \mbox{III}, which is not clearly discernible in the data feature structure, corresponds to \textbf{P10}. 

We also use this synthetic dataset to assess the impact of distance penalty weight $\lambda$. Focusing on the solely class probability-based projection, fixing $\alpha$ to 1, we performed multiple runs of our algorithm with varying $\lambda$ values and random initialization. 
The results are presented in \cref{fig:lambda}. Our findings reveal that the distance penalty is inadequately enforced when $\lambda$ is too small, such as $\lambda<0.1$. Consequently, some data points shoot far away, making the result challenging to interpret.
Conversely, when $\lambda$ exceeds a certain threshold, such as when $\lambda>8$, the distance penalty becomes dominant, causing all data points to be drawn towards a single line. Overall, there is a broad range of values for which $\lambda$ gives good results. However, the threshold from which dramatic change happens can differ for different class probability data. We suggest a default setting between 0.1 to 0.5, which yields satisfactory results in our experiments. 

\subsection{Classifier Analysis}

In this experiment, we present the results of using the Fashion MNIST dataset~\cite{fashionmnist} to test our method and demonstrate its application in classifier analysis. The Fashion MNIST dataset is composed of grayscale images of 28x28 pixels. For this study, a total of 4000 images were selected, ensuring an equal distribution among ten classes, including T-shirts, trousers, pullovers, etc. We further split the selected images into 2800 training and 1200 test samples. A convolutional neural network (CNN) classifier was trained using the training samples, achieving an accuracy of 85\% on the test samples. \textcolor{myred}{To obtain class probabilities that better reflect the inherent uncertainty, we apply Monte Carlo dropout (MC-Dropout) method~\cite{mcdropout} into our CNN model.} Then we applied class-constrained t-SNE to all 4000 samples, using both probability-based classification results and pixel-based feature values. The DR outputs with three $\alpha$ values are shown in \cref{fig:cases}a (top).

\begin{figure}[b]
  \centering 
  \vspace{-0.3cm}
  \includegraphics[width=\linewidth]{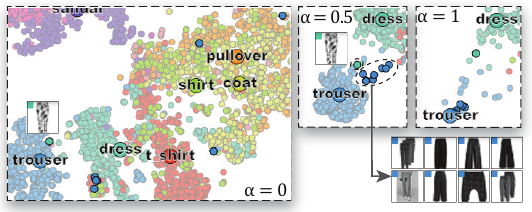}\vspace{-0.3cm}
  \caption{%
     Instances whose data feature similarities and model results conflict are pushed towards the intermediate regions between the main clusters using our method when $\alpha = 0.5$.
  }
  \label{fig:fashioncase}
  \end{figure}

It is observed that the class probability structure ($\alpha=1$) is roughly consistent with the data feature structure ($\alpha=0$), indicated by similar layouts of class landmarks and the degree of separation between class clusters in both structures.
When $\alpha=0.5$, two structures equally contribute to the combined projection result, which leads to enhanced visual separation among the classes of \enquote{shirt,} \enquote{coat,} \enquote{t-shirt,} \enquote{dress,} and \enquote{pullover} while mostly preserving the inner structure \textcolor{myred}{(\textbf{P8})}.
Furthermore, instances that exhibit conflict between data feature similarities and model results are pushed out of the main clusters. For example, the circled blue points in \cref{fig:fashioncase} are not located in the main cluster of class \enquote{trouser} \textcolor{myred}{(\textbf{P9})}. When changing $\alpha$ to 0, we observe that these data points are distributed dispersedly in the data feature structure, indicating that they are not representative of class \enquote{trouser} in terms of feature values. However, these points have high probabilities of belonging to the class \enquote{trouser,} as shown when $\alpha=1$. Further checking of the original images confirms that the model correctly classifies these instances.
Furthermore, the circled green point is classified as \enquote{dress,} but it is actually of class \enquote{trouser,} which accords with its data feature similarity \textcolor{myred}{(\textbf{P9})}. From the above example, we can see that examining instances whose data feature similarities and model results largely conflict is beneficial to examine model performance and identify problematic cases. However, identifying these data points from a single perspective projection or two perspective projections side by side as presented in Embedding Comparator~\cite{embeddingcomparator}, can be challenging due to their dispersion in the data feature structure or concealment among highly certain data points in the class probability structure. The combination presented by our method pulls out these instances from the main clusters and makes them easier to be identified. With the help of animation between projection results of different $\alpha$ values, users can understand how data features and class probabilities relate to each other. 

\subsection{Document Topic Analysis}

The second dataset we use to illustrate our method is the IEEE VIS publication dataset~\cite{vispapers}. We extracted the abstracts and keywords from 3430 publications and removed frequently occurring words such as \enquote{visualization} and \enquote{analysis.} Next, we applied Latent Dirichlet Allocation~\cite{LDA} to extract eight topics and calculate the probability of each paper belonging to each topic. To transform the textual data into a computable format suitable for measuring paper similarity, we used doc2vec~\cite{doc2vec} to generate 100D vectors. These vectors, along with the topic probabilities, were then used as inputs for class-constrained t-SNE. As depicted in \cref{fig:cases}b (top), $\alpha$ was gradually increased to achieve a smooth transformation of DR outputs from displaying data feature structure ($\alpha=0$) to class/topic probability structure ($\alpha=1$).

\begin{figure}[tb]
  \centering 
  \includegraphics[width=0.86\linewidth]{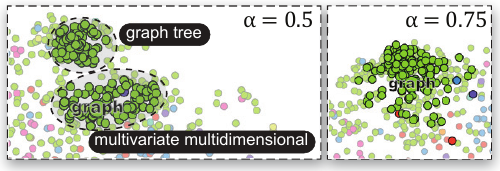}\vspace{-0.3cm}
  \caption{%
    Two clusters with the same topic but different data feature characteristics remain when $\alpha = 0.5$ and merge when $\alpha = 0.75$.
  }
  \label{fig:topiccase}\vspace{-0.4cm}
  \end{figure}

Unlike the data feature structure of the Fashion MNIST dataset manifested as clearly separated visual clusters, the projection result of the 100D document embeddings implies a relatively continuous manifold rather than discrete clusters. The topic probability structure also suggests it with a big blob meaning that the data is difficult to partition into distinct topics. However, from the class landmark and data point layout in both structures, we can still see a separation between infovis papers with topics of keywords - \enquote{graph,} \enquote{model,} \enquote{document,} and \enquote{uncertainty} and scivis papers with topics of keywords - \enquote{field,} \enquote{flow,} \enquote{volume,} and \enquote{surface.} When combining these two structures with $\alpha=0.5$, although there are no clear borderlines between topic clusters \textcolor{myred}{(\textbf{P8})}, we can infer the coherence of topics by observing the coherence of topic clusters in the projection view. Specifically, blue-colored and red-colored data points with topics of \enquote{uncertainty} and \enquote{document} are relatively scattered and mixed with pink-colored and green-colored data points with topics of \enquote{model} and \enquote{graph,} implying potentially low coherence of them. An interesting finding is two separate clusters for the topic of \enquote{graph} as shown in \cref{fig:topiccase} when $\alpha=0.5$ \textcolor{myred}{(\textbf{P2})}. Examination of the corresponding papers in them reveals that papers in one cluster are more about \enquote{graph} and \enquote{tree,} while the other one is associated with \enquote{multivariate} and \enquote{multidimensional.} Splitting this topic into two sub-topics may provide a more coherent representation of the underlying concepts. Note that when increasing $\alpha$ to 0.75, the inner clusters of this topic are not preserved anymore. 
\textcolor{myred}{
\section{Comparative Evaluation}\label{sec:comparison}
To further validate our method, we conduct a comparative analysis with the naive method as baseline presented in \cref{sec: referencemethod} and evaluate them with projection quality metrics. The projection results using the baseline method are displayed in \cref{fig:syntheticalpha}
 (bottom) and \cref{fig:cases} (bottom).}
 
\textcolor{myred}{
\textbf{Quality Metrics.}
To evaluate the projection results, we select three quality metrics, i.e., trustworthiness ($M_{t}$), continuity ($M_{c}$), and Class Consistency Measure (CCM). $M_{t}$ and $M_{c}$ have values in [0,1], with 1 being the best. $M_{t}$~\cite{trustworthiness} evaluates whether $k$ nearest neighbors of points in the projection space are also close in the original space. Conversely, $M_{c}$~\cite{trustworthiness} measures whether the neighborhoods of $k$ points in the original space are maintained in the projection. Since our method uses position proximity between data points to only express data feature similarity, we use $M_{t}$ and $M_{c}$ to evaluate the preservation of data feature structure when changing $\alpha$. In line with previous study~\cite{metrics}, we choose $k=7$ for our evaluation. CCM~\cite{CCM} (also called distance consistency (DSC)~\cite{DSC}) assesses the ability of the projection to faithfully convey the class structure, with values in [0,1] and 0 being the best. From the perspective of human perception of class separation, it is expected that the distance from a point to its corresponding class centroid should be minimal compared to distances to other class centroids. CCM quantifies the proportion of points that violates this centroid distance measure. Although the class structure measured by CCM is not identical to the class probability structure our method tries to preserve, CCM is demonstrated as an effective measure for visual separation of classes in scatterplots~\cite{metricevaluation}.}

\begin{figure}[tb]
  \centering 
  \includegraphics[width=\linewidth]{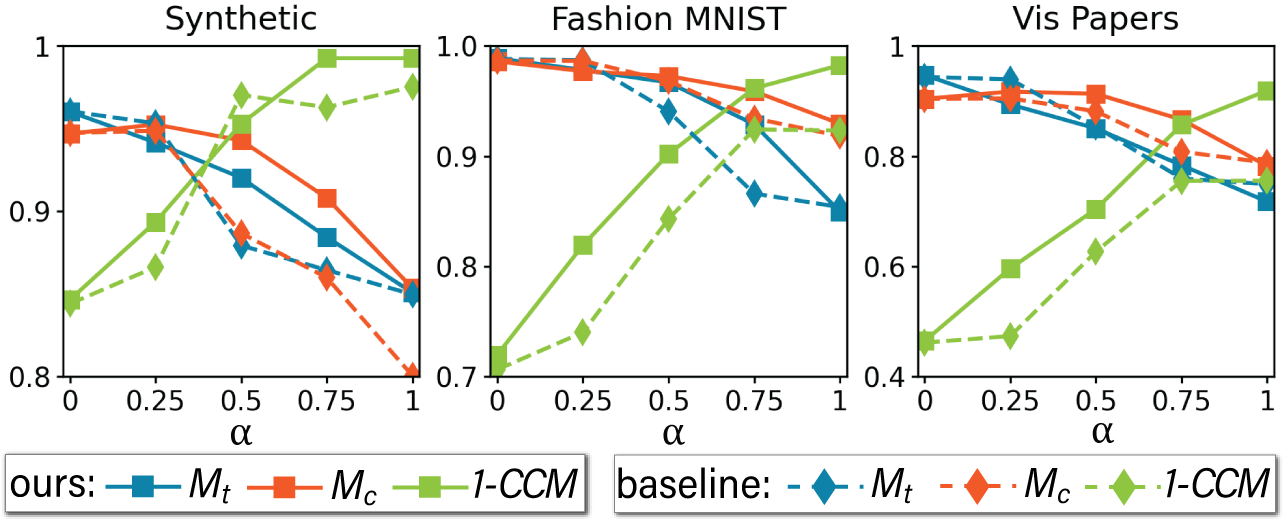}\vspace{-0.2cm}
  \caption{%
    Quality metrics of our method and the baseline method.
  }
  \label{fig:qualitymetrics}\vspace{-0.4cm}
  \end{figure}

\textcolor{myred}{
\textbf{Results.} By analyzing the DR results as shown in \cref{fig:syntheticalpha} and \cref{fig:cases}, we can conclude by comparing our method with the baseline method. First, the global class relationship in our method is reflected by the proximity between class landmarks and is enhanced with increased $\alpha$. As shown in \cref{fig:syntheticalpha} (top), when $\alpha=1$, the class landmark of class 3 is farther to class 1 than class 2, implying classes 1 and 3 are more distinguishable than classes 1 and 2. However, in \cref{fig:syntheticalpha} (bottom), the green points of class 3 are closer to the yellow points of class 1 than the red points of class 2 since global class information is not preserved in the baseline method. Second, as stated before, the class probability relationship between instances and classes is indicated by the proximity between data points and class landmarks, while the baseline method only considers the relationship between data points. Last, the data points whose data feature and class information conflict are more discernible when two structures are combined using our method. When $\alpha=0.5$, outliers \mbox{II} and \mbox{III} in \cref{fig:syntheticalpha} (top) are indiscernible in \cref{fig:syntheticalpha} (bottom). Similar properties can be seen in \cref{fig:cases}. Quality metric results for the three datasets are displayed in \cref{fig:qualitymetrics}. When mixing the data feature structure with the class probability structure, our method generally preserves the data feature structure better, implied by higher $M_{t}$ and $M_{c}$. As expected, our method shows better visual separation with $\alpha$ increasing. }

 \section{Use Scenario}\label{sec:usescenario}
In this use scenario, we will show how class-constrained t-SNE produces DR results which can serve as a guide for annotators in selecting instances to label during the visual-interactive labeling (VIL) process.

VIL refers to the process in which human annotators manually select and assign labels to data in a collaborative, iterative manner with the help of a visual interface~\cite{VIL}. Typically data is projected into 2D space and visualized as a scatterplot for user interaction. During the labeling process, a classification model is trained and updated iteratively based on the newly labeled data, the classification results of which can serve as guidance for instance selection. Once the model is well-trained, it can be utilized to generate labels for unlabeled or new data. Bernard et al.~\cite{labelingoptimal} present a ULoP (Upper Limit of Performance) strategy for instance selection which is demonstrated as a quasi-optimal strategy for labeling data. They generalize this ULoP strategy into three phases. The initial phase focuses on capturing cluster structures and obtaining a uniform sampling of the feature space. The following phase aims to capture the coarse shapes of all classes. The final phase deals with the refinement of poorly-separated classes and outliers. This whole process aligns with the broader conversion from selecting instances based on data feature structure when no data is labeled and the model is untrained to gradually refining the model more based on the model result. If this transformation is directly reflected by the projection result, users can easily select instances while adhering to this strategy. Class-constrained t-SNE has the potential to support such transformation.


\begin{figure*}[tb]
  \centering 
  \includegraphics[width=\linewidth]{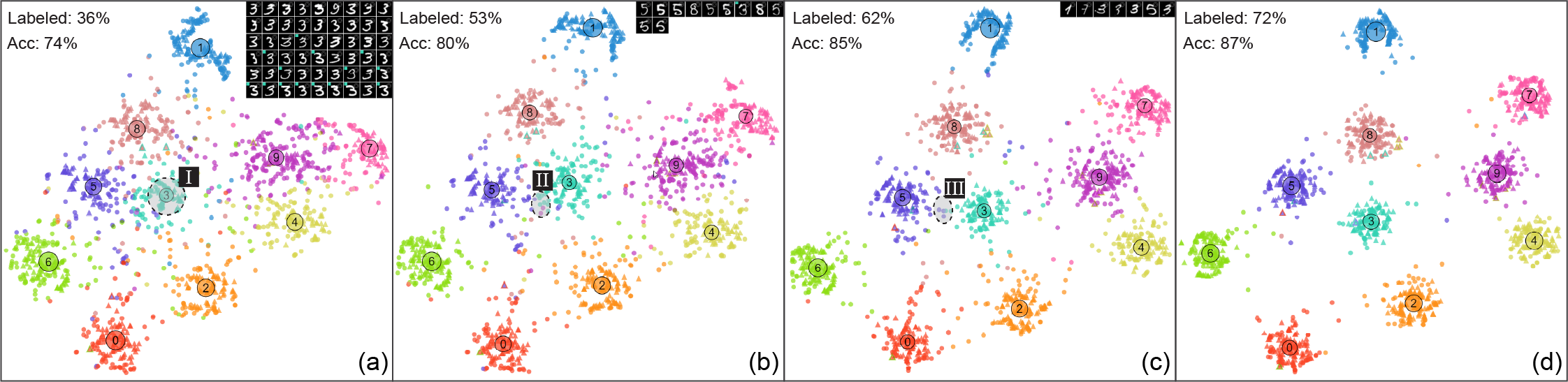}\vspace{-0.1cm}
  \caption{%
    Projection results of MNIST dataset after each model update during the VIL process using class-constrained t-SNE. Images correspond to the selected data points in the circle. The size of class landmarks encodes the number of labeled instances for the corresponding classes.
  }
  \label{fig:labelprogress}\vspace{-0.3cm}
  \end{figure*}

For this use scenario, we developed a visual interface to support human labeling. To make the labeling process manageable, we use the easy-to-understand MNIST data set representing handwritten 0-9 digits, commonly used in VIL-related studies. 2000 images are designated for labeling and model training, and 1000 images are allocated for evaluating model accuracy. Each of the ten classes is equally represented in training and test data. A multi-layer perceptron of one hidden layer is employed for incremental classification. Upon reaching a point where annotators deem they have adequately labeled the data based on the visual patterns apparent in the current DR result, the newly labeled data will be used to update the model. The class probabilities \textcolor{myred}{-- softmax outputs of the updated model -- }will then be utilized in the class-constrained t-SNE to update the DR result, which yields new visual patterns for the next iteration of labeling. In order to accomplish the gradual transformation of DR outcomes from data feature structure to class probability structure, we set $\alpha=(test\_acc)^{2}$. When the model is not well trained, its result is unreliable, reflected by the low test accuracy. 
In our setting, images with identical classes can be labeled with one labeling operation if selected rather than labeling them individually. Thus it is preferable that images with very similar feature values, which highly likely belong to the same class, are clustered together in the resulting DR to facilitate selecting them all together. This is why we enlarge the weight of the data feature structure by using $\alpha=(test\_acc)^{2}$ rather than $\alpha=test\_acc$.

Initially, no instance is labeled, and the model is untrained. Then $\alpha$ is set to 0, thereby solely preserving the data feature structure in the DR result. 
After preliminary labeling based on visual clusters in the data feature structure, the labeled data is utilized for training the classifier, resulting in a 74\% accuracy on the test dataset. The class probabilities derived from this classifier are then utilized to generate a DR result with class-constrained t-SNE. From the dispersion of classes in the resulting projection (\cref{fig:labelprogress}a), one can see that data points of different classes do not generate clear clusters. To enhance the coarse shapes of classes, we select the central part of each class -- the instances around the class landmarks -- for labeling, such as \cref{fig:labelprogress}\mbox{I}. These instances are regarded as representative of class 3 by the model. Confirming their labels facilitates the classifier's ability to capture the coarse shapes of this class. Following labeling all the classes' central parts, we update the model based on newly labeled instances, which results in 80\% test accuracy.
As seen in \cref{fig:labelprogress}b, clusters of different classes become more condensed, although most class clusters are not well-separated. To refine these poorly-separated classes, we select and label the data points at the boundary of the class clusters, e.g., \cref{fig:labelprogress}\mbox{II}. After this iteration, the test accuracy increases to 85\%. The class clusters are refined to exhibit clear separation, as displayed in \cref{fig:labelprogress}c. We can further refine class separation by labeling instances at the boundary of class clusters as shown in \cref{fig:labelprogress}\mbox{III}. Besides, there are mainly outliers scattered in the projection result, representing instances that are difficult for the model to classify or have distinct data feature values. After another two iterations of labeling, we fine-tune the classifier with the newly labeled instances, resulting in an improved test accuracy of 87\%. The class separations in the resulting DR result (\cref{fig:labelprogress}d) are further enhanced based on the new class probabilities. Subsequent fine-tuning should focus on labeling outliers, as well as the unlabeled instances that are close to misclassified instances, whose labeling may help correct the wrong classification since they are similar in the data feature values.

In this VIL scenario, our method generates DR results based on class probabilities of the incrementally updated model, which directly reflect model evolution and show visual patterns that facilitate instance selection and labeling in different phases following the ULoP strategy. 
This approach transforms the labeling process into a visual objective, with the aim of enhancing the differentiation of class clusters.

\section{Discussion}\label{discussion}
This section presents the limitations of our method and proposes avenues for further investigation, encompassing considerations about domain, algorithm, representation, and application. 

\textcolor{myred}{\textbf{Domain.} In our method, the class probabilities of each instance always sum up to one, ensuring the explainability of class information based on the position proximity between data points and class landmarks. This is the primary assumption for our class probability input. Therefore, softmax outputs used in \cref{sec:usescenario} are qualified as input of our method, although it tends to overestimate model certainty. Probabilistic models or calibration methods are recommended if more reliable uncertainty estimates are required. However, in the context of multi-label classification, the probabilities assigned for a single instance may not sum up to one. This deviation from our assumption challenges the ability to interpret the positions of data points in relation to the class probabilities.
As a result, our method is not suitable for multi-label datasets. Moreover, our current implementation does not directly support a semi-supervised setting, which
requires a nontrivial extension to ensure a balanced treatment of data points with or without class probabilities. As this is not the focus of our work, we leave it as future work to adapt our method for a semi-supervised setting. }

\textbf{Algorithm.} 
Our method initializes the optimization process with the previous projection result to achieve a smooth visual transition between continuous and varying perspectives. However, different initializations may impact the optimization process and result in suboptimal projection results. Nevertheless, a balance has to be achieved between possibly suboptimal results and seamless animation to preserve the mental map and aid in the continual perspective-based tracking of instances. Therefore, seamless animation may compensate for the potential shortcomings of suboptimal outcomes. Furthermore, users can always initiate the projection with a random initial position if they hold a particular weighted perspective of interest. 
\textcolor{myred}{In the current implementation, 2D embeddings are initialized randomly from Gaussian distributions. There are alternative options to initialize data points, e.g., using PCA~\cite{PCAInitialization}, which we did not explore in this study. 
Additionally, our proposed method for integrating class probability information has the potential for adaptation to other DR techniques that employ an explicit loss function and necessitate iterative optimization of data points, e.g., UMAP~\cite{UMAP}. However, considering the variations in loss functions employed, how to adapt it to different DR methods requires further exploration, which we leave as future research. }

\textbf{Representation.} Introducing class landmarks in the DR process enables displaying the relationship between instances and classes. These class landmarks as iconic representations of classes or other items can be further encoded to enrich information related to these entities. For example, to display the distribution of keywords associated with each topic in topic modeling, class landmarks can be replaced by bar charts.

\textbf{Application.} This paper mainly presents a method to combine data features and class probabilities. We illustrate it through experiments that mainly focus on classification or topic analysis. However, our method can potentially extend to be applied to other domains. For example, recommender systems model the likelihood of users interacting with particular items based on their past behavior. Our method can aid such analysis by allowing the examination of user similarity and the relationship between users and items.

\section{Conclusion}

In this paper, we propose class-constrained t-SNE, a DR-based visualization method that combines the data feature structure and class probability structure in one projection view. Users can control the balance between the two structures. Class-constrained t-SNE also facilitates the comparison of the two structures by providing a smooth transformation from one structure to the other. The experiments illustrate that our approach improves the interpretability of the projection results from the class viewpoint. Moreover, generated visual patterns can facilitate classifier and document topic analysis, where data features and class probabilities are generally analyzed together. \textcolor{myred}{A comparative analysis with a baseline method further demonstrates the superiority of our method.} Additionally, we present a VIL use scenario to illustrate its utility. Our current class-constrained t-SNE implementation is an adaptation of traditional t-SNE, \textcolor{myred}{which is available from \href{https://github.com/alicelh/class-constrained-t-SNE}{https://github.com/alicelh/class-constrained-t-SNE}}. Future work should consider accelerating our method by integrating it with accelerated variants of t-SNE, such as Barnes-Hut t-SNE~\cite{BHt-SNE}.

	

\bibliographystyle{abbrv-doi-hyperref}

\bibliography{template}


\appendix
\section{Experiment Results}

\begin{figure*}[tbh]
  \centering 
  \includegraphics[width=\linewidth]{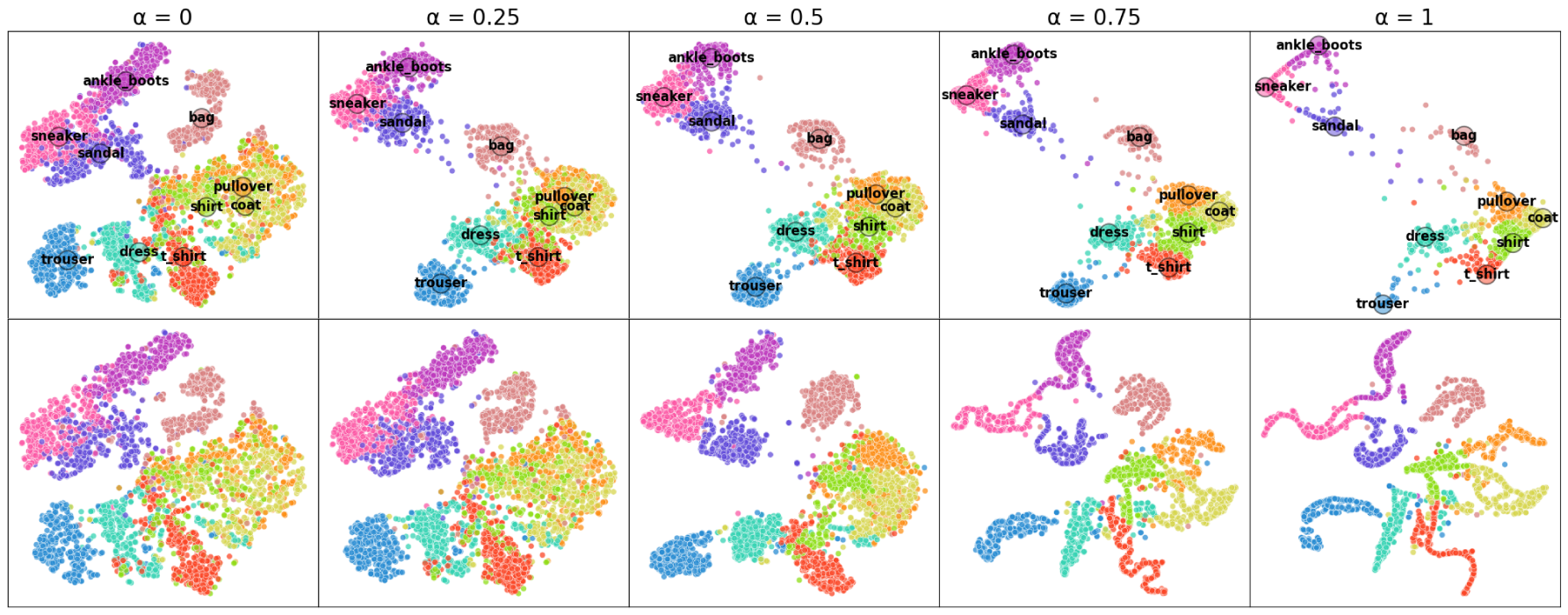}\vspace{-0.2cm}
  \caption{%
    Projection results of the Fashion MNIST with various $\alpha$ values using our method (top) and a baseline method (bottom).
  }
  \label{fig:fashionmnist5}
\end{figure*}
\begin{figure*}[tbh]
  \centering 
  \includegraphics[width=\linewidth]{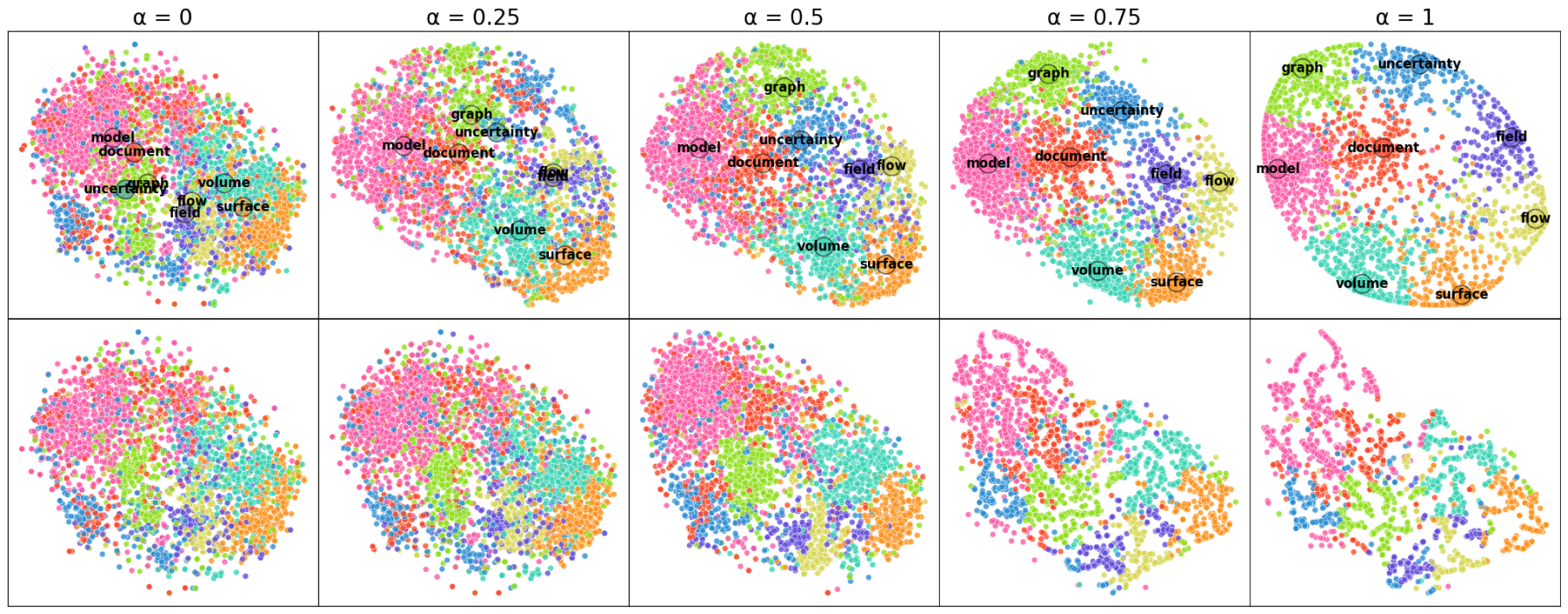}\vspace{-0.2cm}
  \caption{%
    Projection results of the VIS publication dataset with various $\alpha$ values using our method (top) and a baseline method (bottom).
  }
  \label{fig:vispapers5}
\end{figure*}

The projection results of the Fashion MNIST and VIS publication datasets with five continuous and varying $\alpha$ values using our method and the baseline method are shown in \cref{fig:fashionmnist5} and \cref{fig:vispapers5}.

\section{Visual labeling interface}
\begin{figure*}[tbh]
  \centering 
  \frame{\includegraphics[width=0.7\linewidth]{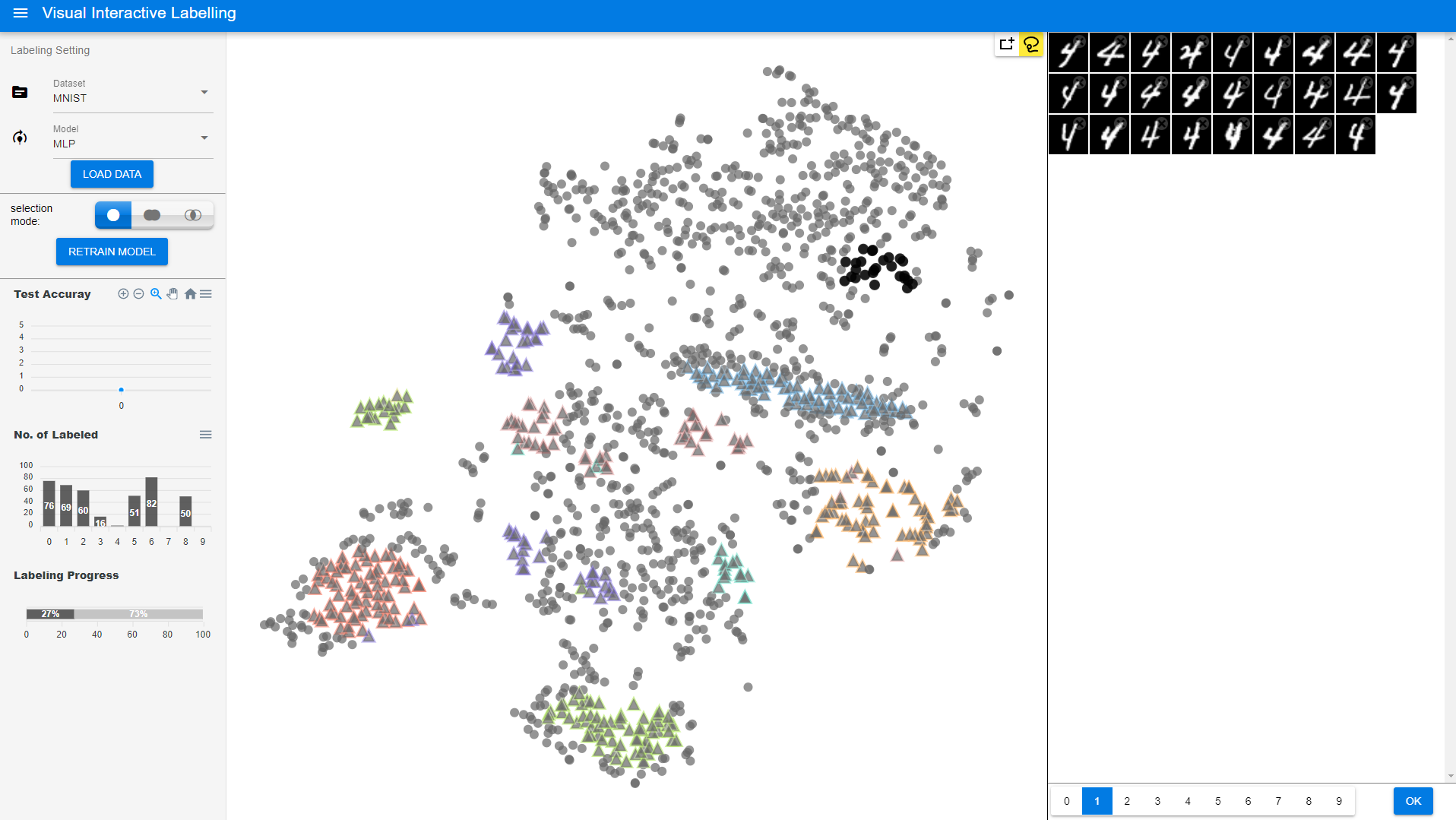}}
  \caption{%
    The visual interface to support human labeling. Unlabeled data points are shown as circles, and labeled data points are shown as triangles. The filling color represents the predicted class, while the border color encodes the labeled class. The filling color is black when the model is untrained at the beginning.
  }
  \label{fig:labelinterface}
  \end{figure*}

We develop a visual interface for supporting human labeling in the visual-interactive labeling (VIL) use scenario as shown in \cref{fig:labelinterface}.

Initially, no instance is labeled, and the model is untrained. Then $\alpha$ is set to 0, thereby solely preserving the data feature structure in the DR result. In this initial phase, adhering to the ULoP (Upper Limit of Performance) strategy, selecting compact clusters for labeling is preferable, wherein instances are more likely to belong to the same class. In the region where data points of different classes are mixed, we select subclusters uniformly to ensure that the selected data points equally spread in the DR results, as exemplified in \cref{fig:labelinterface} where a small subcluster of class 4 is selected and labeled.

\end{document}